\documentclass[aps,prb,amsmath,amssymb,superscriptaddress, showpacs, twoside]{revtex4}

\input {epsf.sty}
\usepackage{graphicx}% Include figure file
\newcommand\ie {{\it i.e. }}
\newcommand\eg {{\it e.g. }}

\newcommand\etal{{\it et.al. }}
\newcommand\noi{\noindent}
\newcommand{\ee}{\end{eqnarray}}
\newcommand{\pref}[1]{(\ref{#1})}

\newcommand\ket [1] {|#1 \rangle }
\newcommand\bra [1] {\langle #1 |}

\newcommand{\av}[1]{\langle #1\rangle}
\newcommand{\tr}{\mathrm{Tr}}

\newcommand\half{\frac 1 2 }

\newcommand{\be}[1]{\begin{eqnarray} \mbox{$\label{#1}$} }
      
\begin{document}
\title{Solitons and Quasielectrons in the Quantum Hall Matrix Model}
\author{T. H. Hansson, J. Kailasvuori, A. Karlhede}
\affiliation{Fysikum, Stockholm University,
AlbaNova University Center,
SE - 106 91 Stockholm, Sweden }

\author{ R. von Unge}
\affiliation{Dept. of Theoretical Physics, Masaryk University, 611 37 Brno,
Czech Republic}

\date{\today}

\begin{abstract}
We show how to incorporate fractionally charged quasielectrons in the finite quantum Hall matrix model. 
The quasielectrons emerge as combinations of BPS solitons and quasiholes in a finite 
matrix version of the noncommutative $\phi^4$ theory coupled to a noncommutative  Chern-Simons gauge field. We also discuss how to properly define the charge density in the classical matrix model, and calculate density profiles
for droplets, quasiholes and quasielectrons. 
\end{abstract}


\maketitle

%  SPECIFIC MACROS  %%
\newcommand{\ok}{\bar k}
\newcommand{\oz}{\bar z}
\newcommand{\rb}{\bar \rho}
\newcommand{\zd}{Z^{\dagger}}

\newcommand\kp{\ell^{2}(\vec k \times \vec p)}
\newcommand\kph{\frac {\ell^{2}} 2 (\vec k \times \vec p)}

\newcommand{\cD}{\mathrm{D}}

\newcommand{\vx} {\vec x}
\newcommand{\vxi}[1]{\vec x_{#1} }
\newcommand{\vk} {\vec k}
\newcommand{\vp} {\vec p}
\newcommand{\vi}[2]{\vec #1_{#2} }

\newcommand{\rhot}{\tilde\rho}

%%%%%%%%%%%%%%%%%%%%%%%%%%%%%%%%%%%%%%%%%%%%%%%%%%%%%%%%%%%%%%%%%%%%%%%%%%%%%

 \section {Introduction}
 During the last few years a new class of models for the fractional quantum Hall (QH) effect has emerged. 
 The basic construction is due to Susskind\cite{suss1}, who observed that the Laughlin states at filling fraction $\nu =   1/(2k +1)$ are naturally described by a noncommutative Chern-Simons (CS) theory, or equivalently, by an infinite matrix model with the lagrangian,
  \be{mat}
 L_{0} = \frac {eB} 2 {\rm Tr} \{ \left( \dot{ X^{a}}  - i[X^{a},
 \hat a_{0}]_{m} \right) \epsilon_{ab}X^{b} +
 2\theta \hat a_{0}\} \, ,
 \ee
  where $X^{a}(t)$, $a=1,2$,  and $\hat a_{0}(t)$ are hermitian matrices --
 the latter being a  Lagrange multiplier imposing the matrix commutator constraint, 
$
 [X^{1},X^{2}] =    i\theta \, .
$
 The area $\theta$ that enters $L_{0}$ is the
 noncommutativity parameter, and $B$ is the transverse magnetic field. 
 
 As it stands, this model has only a single state, since the solution to the constraint,
 which can only be satisfied by infinite matrices, is 
 unique (up to gauge transformations).
 This reflects that the theory is topological and
 thus has no excitations when defined on an infinite plane.\footnote{
 On a manifold with nontrivial topology one expects, in analogy with the commutative case, a 
 ground state degeneracy in the {\em quantum} matrix model. }

The parameter $\theta$ can be interpreted as an
 area per particle, giving  the unique state a constant density  $\rho = 1/2\pi \theta$.
Modifying the constraint by hand, one finds other solutions correponding to fractionally 
charged quasielectrons and quasiholes\cite{suss1}. 

 In an important development, Polychronakos extended the model by  supplementing
 \pref{mat} with the lagrangian,
 \be{fin}
 L_{b} = \Phi^\dagger (i\partial_0 - \hat a_{0})\Phi \ \ ,
 \ee
 where $\Phi$ is a complex bosonic $N$-vector\cite{poly1}.
The $X^a$'s in
 \pref{mat} are now hermitian $N\times N$ matrices,  and the constraint is  changed to
 \be{ptvang}
 [X^{1},X^{2}] =    i\theta - \frac i {eB} 
 \Phi\Phi^\dagger  =  i\theta(1 - \frac 1 \kappa 
 \Phi\Phi^\dagger) \, ,
 \ee
 where $\kappa = eB\theta$ is the so called level number.
 It is striking that this  finite QH matrix model (QHMM) already at the classical level describes several
 key features of the quantum Hall system:
 \begin{itemize}
 \item
 In the presence of a rotationally invariant  confining potential, the groundstate is a  finite size circular ''droplet'' with a constant bulk  density $\rb$ depending on the level number. 
 \item The excitation spectrum is consistent with that of a QH droplet. In particular there are quasihole states in the bulk and gapless quasielectron - quasihole states at the edge.
 \item In the absence of a potential, there is a set of  degenerate low density states corresponding to single particles in the lowest Landau level,  at well separated  positions in the plane.  
\end{itemize}
In particular note that the presence of quasielectron and quasihole excitations takes this description beyond that of a classical incompressible fluid, and we shall see below how  the model also describes how QH droplets are formed from well separated particles in a strong magnetic field. 

Quantizing \pref{mat}, and assuming the underlying matrix degrees of freedom to be fermionic, Susskind showed 
that the density is quantized at the Laughlin fractions $\nu = \bar\rho/\rho_0 =  1/(2k+1)$, where $k$ is  integer and
$\rho_0 = eB/2\pi$ (when $\hbar=1$) is  the density of states in a single Landau level.\footnote{
In his original paper, Susskind argued, based on the underlying fermion statistics,  that the level number $\kappa$ is quantized as an odd integer so that the classical relation $\kappa = eB\theta$, or $\rho = 1/2\pi\theta = \rho_0/\kappa$, implies that the density is quantized at the Laughlin fractions. This argument was later corrected by Polychronakos\cite{poly1} and by Hellerman and Susskind\cite{hell1}. It turns out that in the {\em quantum} matrix model the relation between density and level number is shifted to $\nu = 1/(1+\kappa)$ because of quantum fluctuations. 
There is however also a correction to the relation between the level number and statistics that was used in reference \onlinecite{suss1}. These effects in fact cancel and leave the original result of Susskind unchanged. A topological 
argument for the quantization of $\kappa$ to any integer was given by Nair and Polychronakos\cite{nair1}. }
At a technical level, it was also shown in reference \onlinecite{poly1} that in the presence of a 
quadratic potential, there is an exact mapping of the QH matrix model onto the Calogero model, 
both in the classical and the quantum case. This mapping yields  explicit expressions for both energy levels and wave functions\cite{hell2}. 

In a previous paper we extended the QHMM model further, by constructing a class of conserved charges and  accompanying currents, thus  allowing for a coupling to an external electromagnetic field\cite{hkk}. 
We then went on  to calculate low momentum  response functions in the classical model, in particular:
 \begin{itemize}
 \item The ground state density, being the response to a constant electric potential, $A_0$.
 \item
 The quantum Hall response $\sigma_H$.
 \item
 The response to a weak and slowly varying external $\vec B$ field.
 \end{itemize}
The results were  all in agreement with the known properties of the Laughlin states. 
 
In spite of these successes there are several basic aspects of QH physics which
are not incorporated in the finite matrix model given by \pref{mat} and \pref{fin}. 
Most significantly: 
\begin{itemize}
\item
There is no unambigous definition of density.
\item
There are no quasielectron solutions.
\item
There is no natural way to introduce spin and/or multilayer degrees of freedom analogous to
the usual description in terms of multi-component CS fields\cite{zee}.
\item
There is no generalization to fractions other than the Laughlin ones.
\end{itemize}
In this paper we address the first two points, in some detail, while the 
third is addressed in another paper\cite{hkk2}.
Before turning to the technicalities, we will give some
general comments to the above list, and also briefly discuss the status
of the quantized QH matrix model. 

In reference \onlinecite{hkk} we constructed a class of conserved currents for 
the classical matrix model. As we will discuss  below, there are three 
natural conditions on the charge density operator: it should be non-negative, 
satisfy the classical version of the sine-algebra characteristic of the lowest Landau level, 
and have the correct limit for particles separated much further than the magnetic length. 
Unfortunately, we have not found any definition that satisfies all these demands. 

The absence of quasielectrons in the noncommuative theories is related
to the existence of a minimal area for each particle. A clue as  to how to get
quasielectrons is given in Susskind's original paper where 
they emerged from an {\em ad.hoc.} change of the constraint. 
In a recent paper, Bak \etal showed how the addition of a noncommutatve
scalar field $\phi$, provides a dynamical version of this mechanism, and
gives a model with soliton solutions with charge density larger than 
$\rho_0$\cite{bak}. In section III we shall construct the corresponding 
finite matrix model. 

The problem of spin, (or pseudospin corresponding to \eg a multilayer index) derives from 
the restricted nature of noncommuative gauge theories -- U(N) is the only allowed 
gauge group\cite{doug}. The standard multi-component CS lagrangians employed 
to describe spin and pseudospin, as well as the general classification of abelian QH liquids 
given by Wen\cite{zee}, are all based on the gauge group $U(1)^k$.  

The problem of finding non-Laughlin states, as already mentioned, is superficially the same as for spin -- there is no 
noncommutative version of the standard multi-component CS theories. 
As we show in reference \onlinecite{hkk2}, however, the spin problem can be addressed by introducing
fermionic degrees of freedom and couple them in a judicious way to the bosonic matrices. We know
of no such construction for generating non-Laughlin states, and the initial hope that the matrix theory 
would provide a new and more powerful framework for the classification of QH liquids has so far
been elusive. 

Thus, turning to quantum theory, there is no matrix model where the density is quantized to other 
fractions than the Laughlin ones (except for the trivial case of direct sums), and in particular there is 
no way to get the experimentally prominent Jain series\cite{jain1}, 
$\nu = n/(2pn\pm1)$). At a technical level the quantized QHMM is hard to handle since the current and density operators are mathematically very complicated objects. This means that  although the quantum states of the model are known via the  mapping from the Calogero model\cite{poly1}, it is not possible to calculate density profiles.  Even for the simplest case of two by two matrices the manipulation of exponentials of matrices with (quantum) noncommuting elements is very difficult. 

We already stressed the pros and cons of the {\em classical} matrix model, and the aim of this paper, and 
reference  \onlinecite{hkk2}, is to extend this model to allow for quasielectrons and spin, and also  to 
find a density operator that can describe quasiparticle and edge profiles consistent with what is known
about the QH system. As we shall see, this endeavor has been rather successful, at least on a qualitative level.
Within an extended finite QH matrix model, we can describe QH droplets, exponentially falling edges, 
quasihole and quasielectron excitations. In reference \onlinecite{hkk2} it is also shown how to incorporate 
spin.

The paper is organized as follows. In the next section we discuss the ambiguities in the definition 
of the density operator and give the arguments in favour of our special choice. We then calculate
density profiles for droplets, and quasiparticles and compare with what is expected from other approaches
such as CS mean field theory, and Laughlin wave functions. In section III we first show how to incorporate
 densities larger than $\bar\rho$ by adding a scalar field to the finite matrix model. The resulting theory 
 has soliton solutions with integer charges, and quasielectrons can be constructed by adding holes on top of these solitons.
 We give explicit expressions for the solutions and calculate the density profiles which are again compared
 with alternative descriptions. In the last section we summarize our results and contrast the classical matrix 
 model approach with the standard classical commutative CS description. Some technical points
 about the positivity of the density operator and possible alternative definitions are given in an appendix.

\section{Particles, droplets and quasiholes} 
In this section we shall study the density profiles of various solutions of the finite classical matrix model.
These solutions were all found by Polycronakos\cite{poly1}, who also determined gross characterizations such
as the radius and mean density of the QH droplet, and the charge of the quasihole.  
To calculate the profiles, we must first give a  definition of the density operator. As stressed in
the introduction, our choice, although not unique, gives profiles in good agreement with
those obtained by other methods.

\subsection{The density operator}
In reference \onlinecite{hkk} we constructed a class of conserved currents for 
the classical matrix model. The general form of the charge and current was given by
 \be{charge}
 \rho(\vec y, t)    &=&  {\rm Tr} [    \hat \delta (y^a  - X^a(t)) ]  \\
  \vec j(\vec y, t) &=&  {\rm Tr}[( \dot{\vec X}
 -i[\vec X, \hat a_{0}]_{m})\hat \delta (y^a - X^a(t))]  \, , \nonumber
 \ee
 where $\hat \delta (y^a  - X^a) $ is a matrix-valued kernel. 
The general form of this kernel follows from symmetry considerations
and current conservation,
 \be{mgker}
 \hat \delta (y^a - X^a) = \int \frac{d^2k}{(2\pi)^2}\, f(k_{a} (y^a -
 X^a),
 \epsilon_{ab}k^{a} (y^b - X^b), \theta k^2) \ \
 \ee
 and a  rather natural guess is
 \be{weyl1}
 \hat \delta_W(y^a  - X^a) = \int \frac{d^2k}{(2\pi)^2}\, e^{ik_a
 (y^a - X^a)} g(\theta k^2) \ \ ,
 \ee
 where $g(0)=1$. (The special case $g(x)=1$ is known as Weyl-ordering.)
 For almost diagonal matrices, appropriate for widely separated particles at 
 positions $\vec x_n = \vec X_{nn}$, 
 this corresponds to $\rho_{\vec k} = g(\theta k^2)\sum_n e^{-i\vec k\cdot\vec x_n}$. 
 Clearly $g$ can be thought of as a formfactor, and for the particular choice
 $g(x) = e^{- (x/d)^2}$ we have gaussian ''blobs'' which are 
 very suggestive of maximally localized one-particle wave-functions in the LLL. With this motivation
 we shall use 
 \be{ourdef}
 \rho_{\vec k} = e^{-\frac{\ell^2k^2}{ 2}} \tr {e^{-i\vec k \cdot \vec X} } \, , \ee
 where the length $d$ was chosen as to match the rms radius  
  $\sqrt{\av{ r^2}} = \sqrt 2 \ell \equiv 
(2\hbar/eB)^{1/2} $ which is the appropriate value 
 for an isolated electron in the LLL. 
 
 Before proceeding to use  the formula \pref{ourdef} to calculate density profiles, we mention some
 problems with this definition.  The first, and most severe, is that $\rho(\vec x)$ is not positive 
 definite on the space of matrices satisfying the constraint \pref{ptvang}.  
 This is not obvious, but can be shown by numerical calculations which also indicate 
 that this is mainly a problem for very small systems, typically 
 $N<10$, and also gets more severe with lower $\nu$. For moderately large N, 
very small violations of positivity is seen in
typical density
profiles such as the "droplet" solution shown in Fig. \ref{f:N50compl}
 for $N=50$. For
extreme cases, such as $N=2$, the violation of positivity is large, as
shown in the appendix. We have not been able to show that the 
definition \pref{ourdef}
gives a positive definite density in the limit $N\rightarrow \infty$, 
although our numerics appears to support this possibility.
  
The situation is less favourable for 
 other ordering prescriptions. So will for instance antiordering, defined by
 \be{antior}
 \hat \delta_{\mathrm{ao}} (z-Z,\oz-Z^{\dagger}) = \int \frac{{\rm
 d}^2k}{(2\pi)^2}\,
 e^{\frac{i\ok}{2} (z-Z)} e^{\frac{i k}{2}(\oz- Z^{\dagger})}  \tilde g(\theta k^2)\ \ ,
 \ee
where again $\tilde g(0)=1$,  give strongly fluctuating profiles, and
large negative values for the density even for rather large $N$.\cite{janik} By going outside the class of density operators that can be written on the form
\pref{charge}, \ie as a trace of a matrix kernel, one can define a positive definite 
density operator with the correct limiting behavior for separated particles. 
This construction, which essentially involves taking the square root of a delta function,
has, however, other shortcomings.  Technical details are given  in the appendix.  

A second problem is that we would expect the Fourier components of the 
quantum mechanical density operator to satisfy the following commutation
relation, 
\be{lowest}
[\rho_{\vec k}, \rho_{\vec p}]_\textrm{QM}  = 2i \sin\left(\kph 
\right)e^{\frac {\ell^{2}} 2 \vec k 
\cdot \vec p } \rho_{\vec k+\vec p} \, ,
\ee
which is the sine algebra pertinent to the density operator projected onto the LLL. 
We have not been able to find any definition of the density that satisfies \pref{lowest} except for
$\theta = 0$, where the anti-ordering is known to be correct.  The claim  in reference \onlinecite{hkk} that
 a particular quantum reordering of \pref{antior} satisfies \pref{lowest} for $N=2$ is erroneous.\footnote{
The error is corrected in arXiv:cond-mat/0304271v2. } Actually, by studying the classical limit we can show that there is no quantum ordering of neither the matrix Weyl ordered nor anti-ordered density operators that satisfies \pref{lowest}.

 For readers familiar with the  string theory literature, the following comment  might be of interest.
In string theory one can show that
 the Weyl ordered expression for the density, corresponding
to $g(\theta k^2)=1$ in  \pref{weyl1}, gives the
density of the lower dimensional RR-charged D-branes.
This follows since \pref{weyl1} is nothing but the Seiberg-Witten map for the noncommutative field strength which
implies that it couples to the Ramond-Ramond forms in precisely the correct way to act as a source of the corresponding RR-charge\cite{sw0}.
In  our case there is no such reason to use Weyl-ordering to  define the density of
particles and we may modify this expression as long as it respects the
symmetries of the problem. Note however, that our choice \pref{ourdef}
coincides with Weyl-ordering for the $k=0$ component corresponding to the total charge.

To summarize, we have no {\em a priori} reason to choose 
\pref{weyl1} rather than \eg antiordering, or in fact any other ordering
 in the general class \pref{mgker}.
Similarly, there is  no theoretical motivation for taking any particular $g(x)$. Instead our choice
\pref{ourdef} is phenomenologically motivated, and its usefulness will be demonstrated in
the rest of this paper.

 \subsection{From particles to droplets}
For the gauge choice $\Phi^\dagger =\sqrt{eB\theta} ( 1 1 \ldots 1) $, the constraint \pref{ptvang} is solved 
by the following matrices\cite{poly1} 
\be{symsolcon}
X_{mn}^1 &=&  x_m\delta_{mn}  \\
 X_{mn}^2 &=&  y_m\delta_{mn}-
\frac{i\theta}{x_m-x_n}(1-\delta_{mn}) \nonumber \, , 
\ee 
where we (arbitrarily) chose to diagonalize the hermitian matrix $X^1$. 
For widely separated $x_i$:s, the off-diagonal terms, that are responsible for 
the  ''$\theta$-repulsion'',  are small, and the diagonal 
elements can be interpreted as  the coordinates of the particles. More generally, we can think
of the (gauge invariant) eigenvalues of the matrices $X_i$ as particle coordinates $x_i$ and $y_i$. 
Note, however, that there is no unambiguous way to pair these eigenvalues to 
position coordinates for the particles. 

Another convenient gauge choice is $\Phi^\dagger =\sqrt{N\kappa}
(0 \ldots 0 1)$ and introducing the dimensionless 
complex coordinates 
$Z\equiv \frac{1}{\sqrt{2\theta}}(X^1+iX^2)$,  the constraint takes the form
\be{conA}
[Z,Z^\dagger]= 1-N|N-1\rangle\langle N-1| \, ,
\ee
where the bra-ket notation refers to an oscillator basis as explained in  \eg reference \onlinecite{doug}. 
In the large $N$ limit, this is the usual ladder operator algebra, and the effect of the boundary field
is only at the ''edge'' of the matrix. It is thus natural to seek a solution for $Z$ similar to the
lowering operator $a$ in the $n$-representation. One finds,
\be{dropsolcon}
Z =\sum_{n=0}^{N-1}\sqrt{n}|n-1\rangle\langle n|\, . 
\ee 
We will refer to this as the droplet solution.  
By a $U(N)$-transformation it can be put on the form  
\pref{symsolcon}, with $y_m=0$ (since the matrix elements in  $Z$ are real) and almost equidistantly spaced $x_m$:s, which are the (gauge invariant) 
 eigenvalues of the hermitian combination $(Z+Z^\dagger)/2$.\footnote{
 These eigenvalues are actually the roots of the  $N$-th order Hermite polynomial as
pointed out by Polychronakos in reference \onlinecite{poly1}. }

%%%%%%%%%%%%%%%%%%%%%%%%%%%XXXXXXXXXXXXXXXXXXXXXXXXXXXXXXXXXXXX

\begin{figure}[h]
\centerline{
\includegraphics[width=6cm]{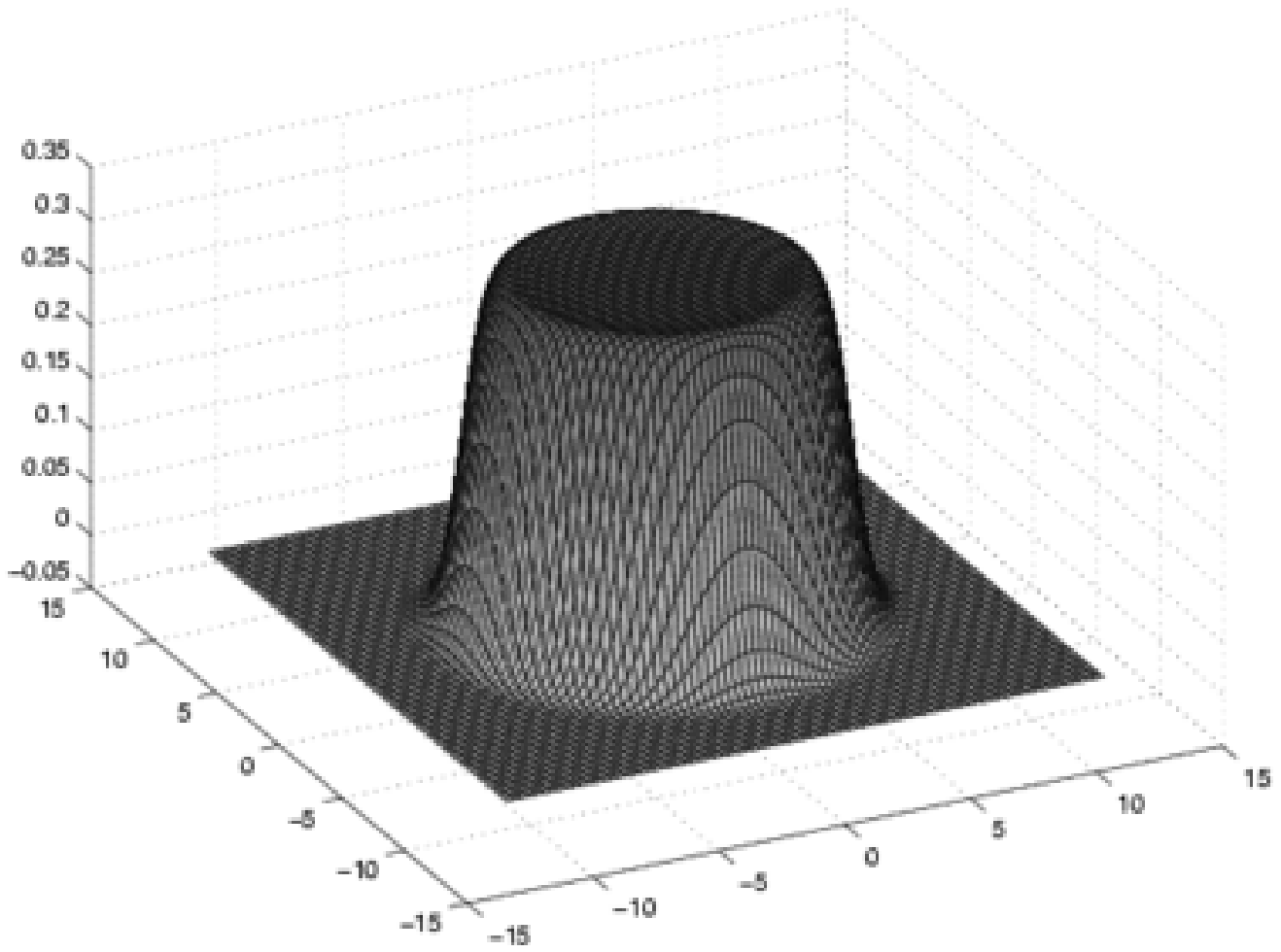}
\includegraphics[width=6cm]{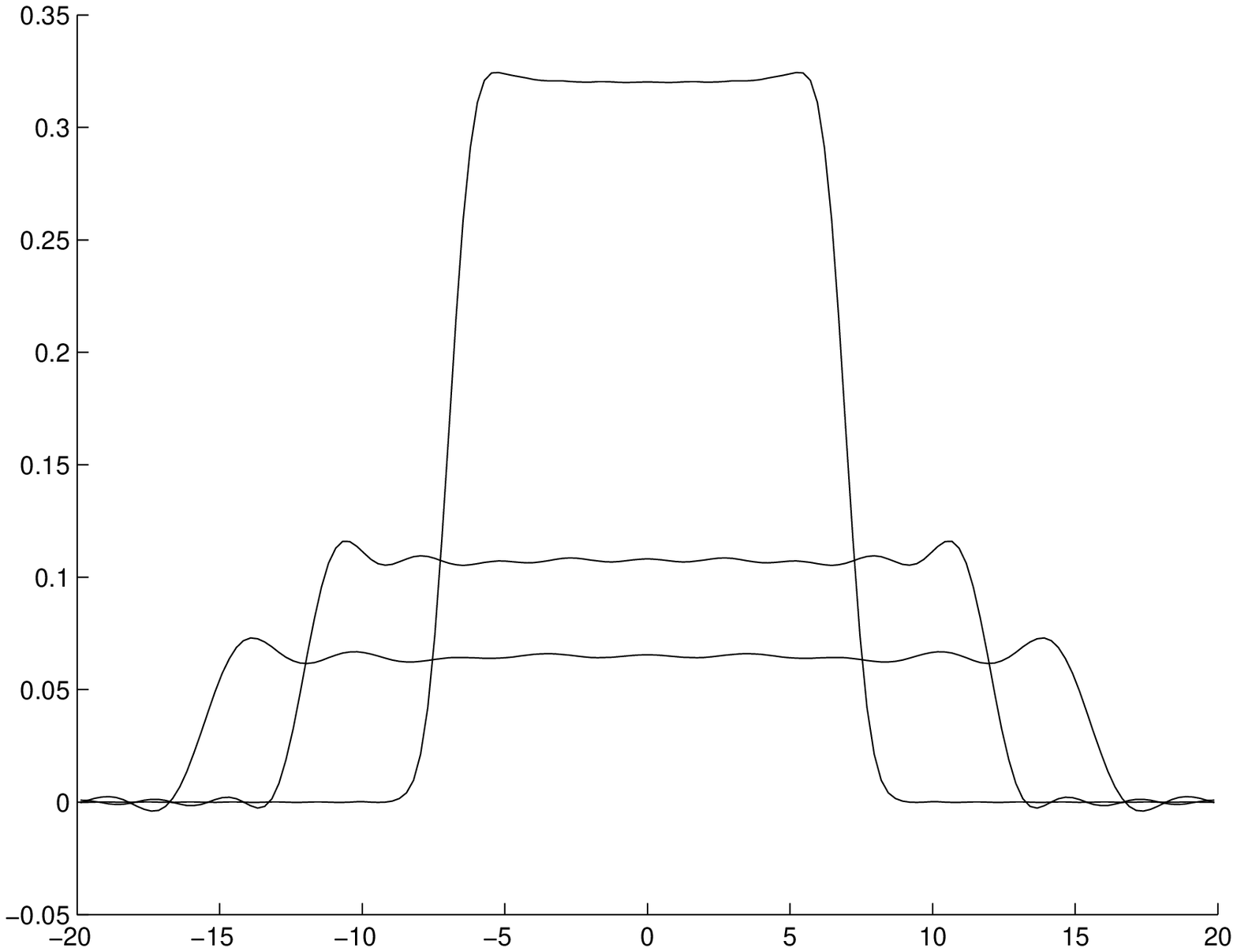}
}
\caption{
 \label{f:N50compl}
To the left  $\rho (\vec{x})$  for  $\nu =1$ as given by the density operator (\ref{ourdef}) and the droplet solution (\ref{dropsolcon}) with $N=50$. We use dimensionless units given by $2\ell^2=2\nu\theta=1$.  To the right  the density $\rho (\vec{x})$ in profile along the $x$-axis is given for $\nu=1,\frac{1}{3},\frac{1}{5}$, respectively. All states have roughly constant bulk densities, but note the increasing wiggles, and the building up of a rim at the edge, for lower $\nu$.  Note also the small violations in positivity at the edge.}
\end{figure}

%%%%%%%%%%%%%%%%%%%%%%%%%%%%%%%%%XXXXXXXXXXXXXXXXXXXXXXXXXXXXX

Using the choice (\ref{ourdef}) for $\rho_{\vec{k}}$ , we can calculate the corresponding $x$-space
density profile $\rho (\vec{x})$, which is shown  in Fig. \ref{f:N50compl}  for $\nu = 1,\frac{1}{3},\frac{1}{5}$ and $N=50$. The lower $\nu$ is, the more pronounced is the up-shooting rim at the edge. Excluding a circular segment containing the rim,  the distribution is very well fitted by the formula
\be{edgefit}
\rho(r,\theta) =\frac { \rho_0} 2  \left(1- \tanh \frac {r - r_0} {\beta \ell} \right)
\ee
with $r_0 \approx 0.99 \sqrt{2\theta N}=0.99\ell \sqrt{2N/\nu}$ for all three $\nu$ and  $\beta \approx 1.05,0.92,0.82$ for $\nu = 1,\frac{1}{3},\frac{1}{5}$ and $N=50$ respectively. This is consistent with the expectation of a constant bulk density and a very rapid 
fall-off at the edge over a distance of the order of the magnetic length. 
\footnote{Taking the logarithm of the density shows that the tail in fact falls faster than the exponential in equation \pref{edgefit}.
A better fit is given by  $e^{-((r-r_0)/\beta \ell)^2}$.  
However, fitting of just the tail is much more uncertain than is a fitting  where also the plateau is used .
}

\begin{figure}
     \includegraphics[scale=0.666]{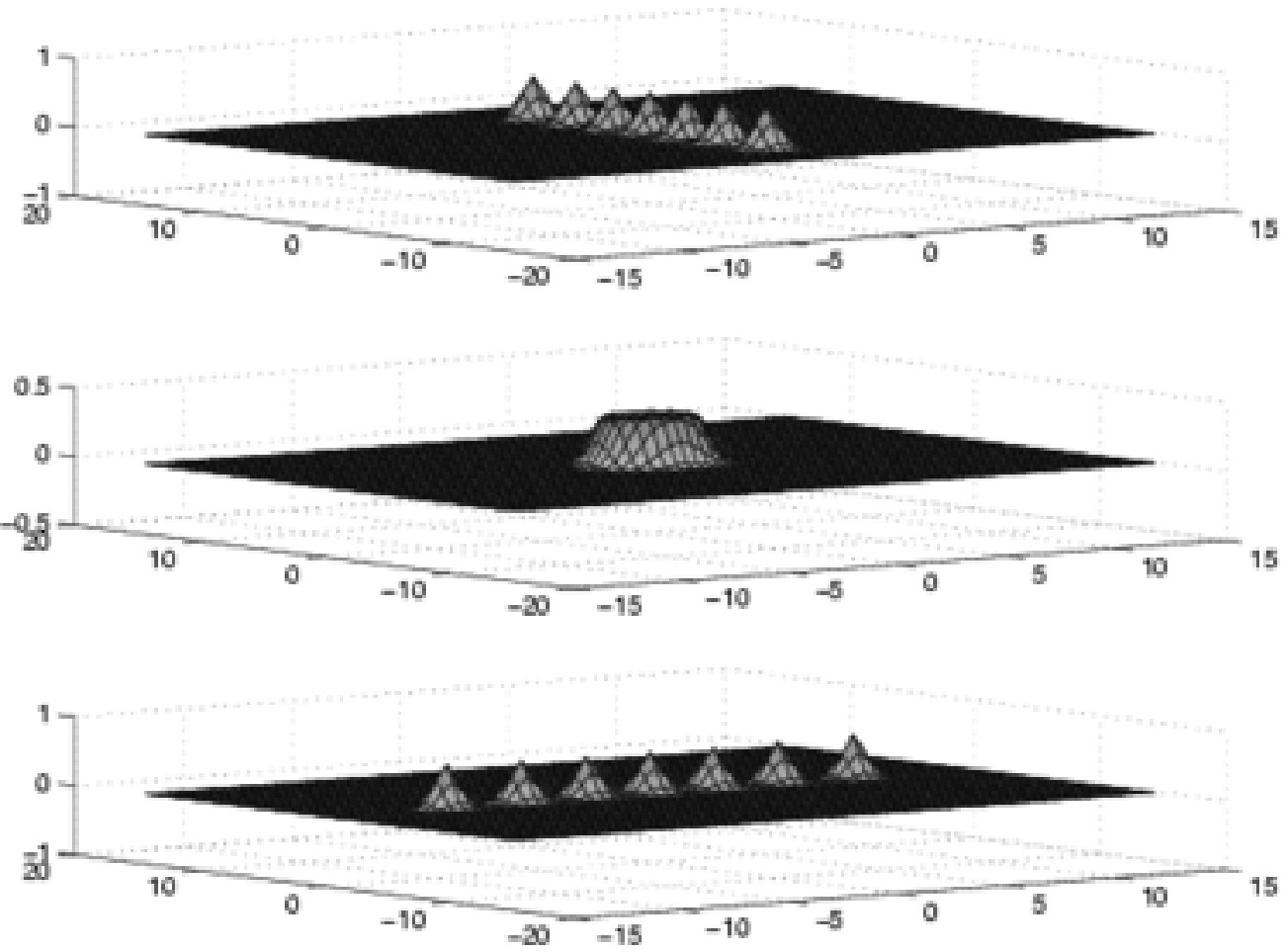}
 \caption{
The compression of seven quantum Hall particles as discusssed in the text.  Note the perfect circular symmetry of the maximally compressed state 
in the middle picture.} 
 \label{f:sem7rhoxyth1}

     \includegraphics[scale=0.666]{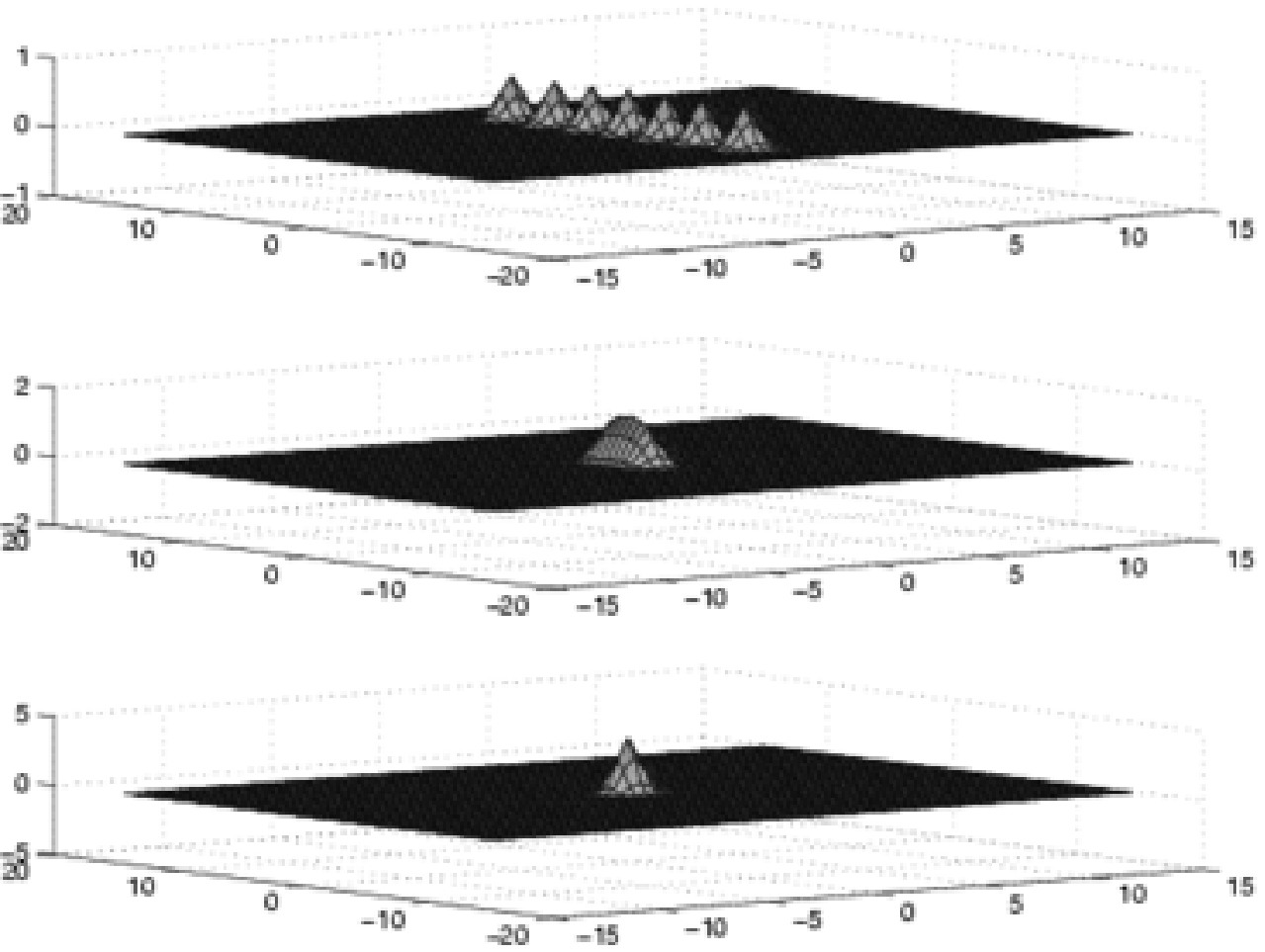}
 \caption{
As in Fig.\ref{f:sem7rhoxyth1} but with $\theta=0$. No droplet is formed and the maximally compressed state in the last picture is just seven superimposed gaussians. }
 \label{f:sem7rhoxyth0}
 \end{figure}

In  Fig. \ref{f:sem7rhoxyth1}  we illustrate how a droplet is formed when several, 
initially  well separated, particles approach each other. The middle figure  is a density
plot of the droplet solution \pref{dropsolcon} for seven particles. From this solution we extracted the eigenvalues
$x_m$, and then generated a set of solutions of the form 
\pref{symsolcon} by scaling the $x_m$:s  by a common factor, $\lambda$. 
The top figure shows the density for $\lambda = 5$, corresponding to particles well separated on the $x$-axis. In the limit of 
large $\lambda$, the $x_m$:s are simply the coordinates of the particles. The bottom picture is for $\lambda = 1/5$. Because of the 
$\theta$-repulsion the particles cannot be compressed further than the droplet, and the result is instead particles separated along
the conjugate $y$-direction. That this effect is entirely due to the finite value of $\theta$ is demonstrated in Fig.  \ref{f:sem7rhoxyth0}, 
which is identical to  Fig. \ref{f:sem7rhoxyth1}, but with the off-diagonal $\theta$-repulsion terms in \pref{symsolcon} 
set to zero (still keeping the same formfactor $g(\theta k^2)$.  In this case no circular droplet is formed and the 
maximally compressed state is simply an overlap of the individual gaussian distributions.

%%%%%%%%%%%%%%%%%%%%%%%%%%%%%%%%%%%XXXXXXXXXXXXXXXXXXXXXXXXXXXXX

%%%%%%%%%%%%%%%%%%%%%%%%%%%%%%%%%%%%%%%%%%XXXXXXXXXXXXXXXXXXXXXXXXXXX
 
 \subsection{The quasihole solution} 
 
\begin{figure}
\centerline{
\includegraphics[width=6cm]{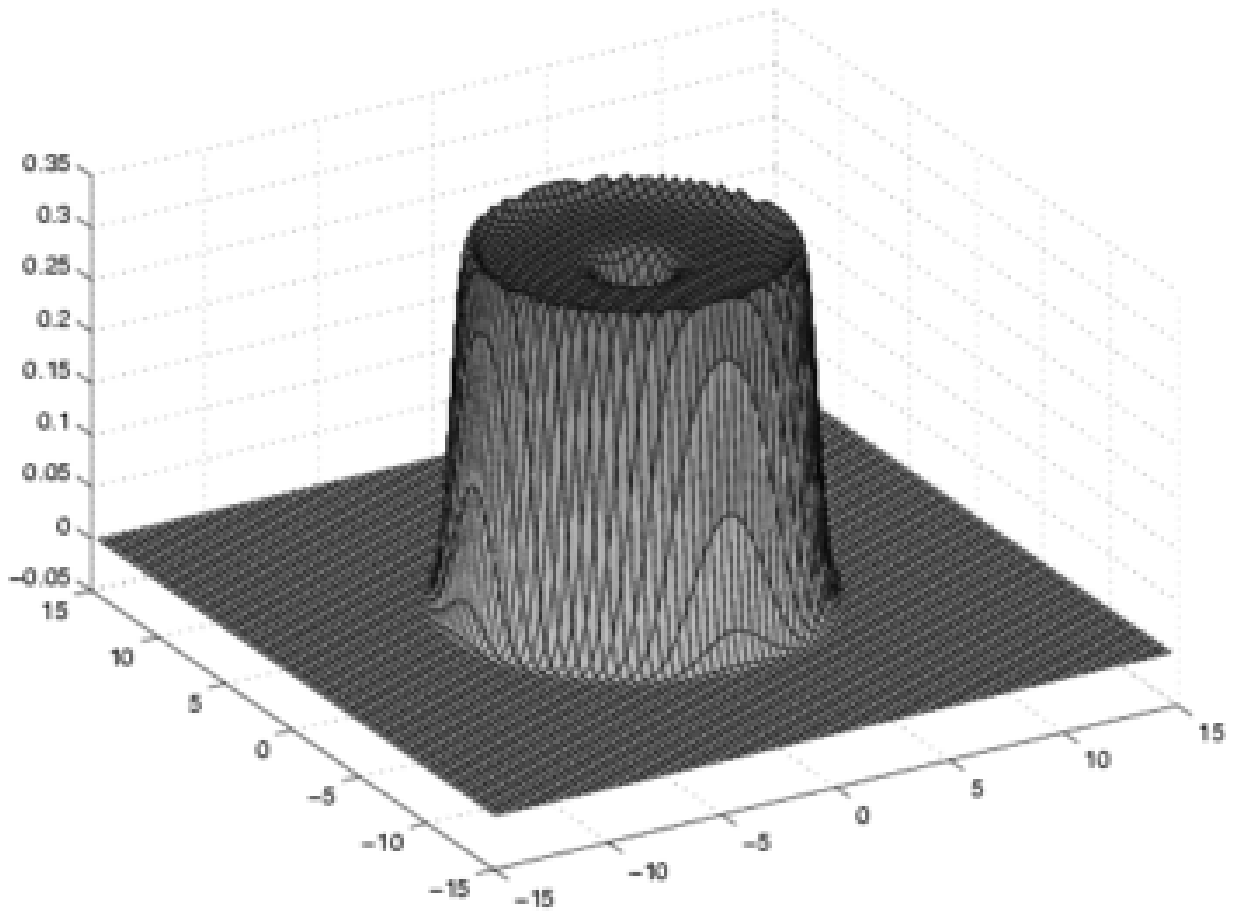}
\includegraphics[width=4.3cm]{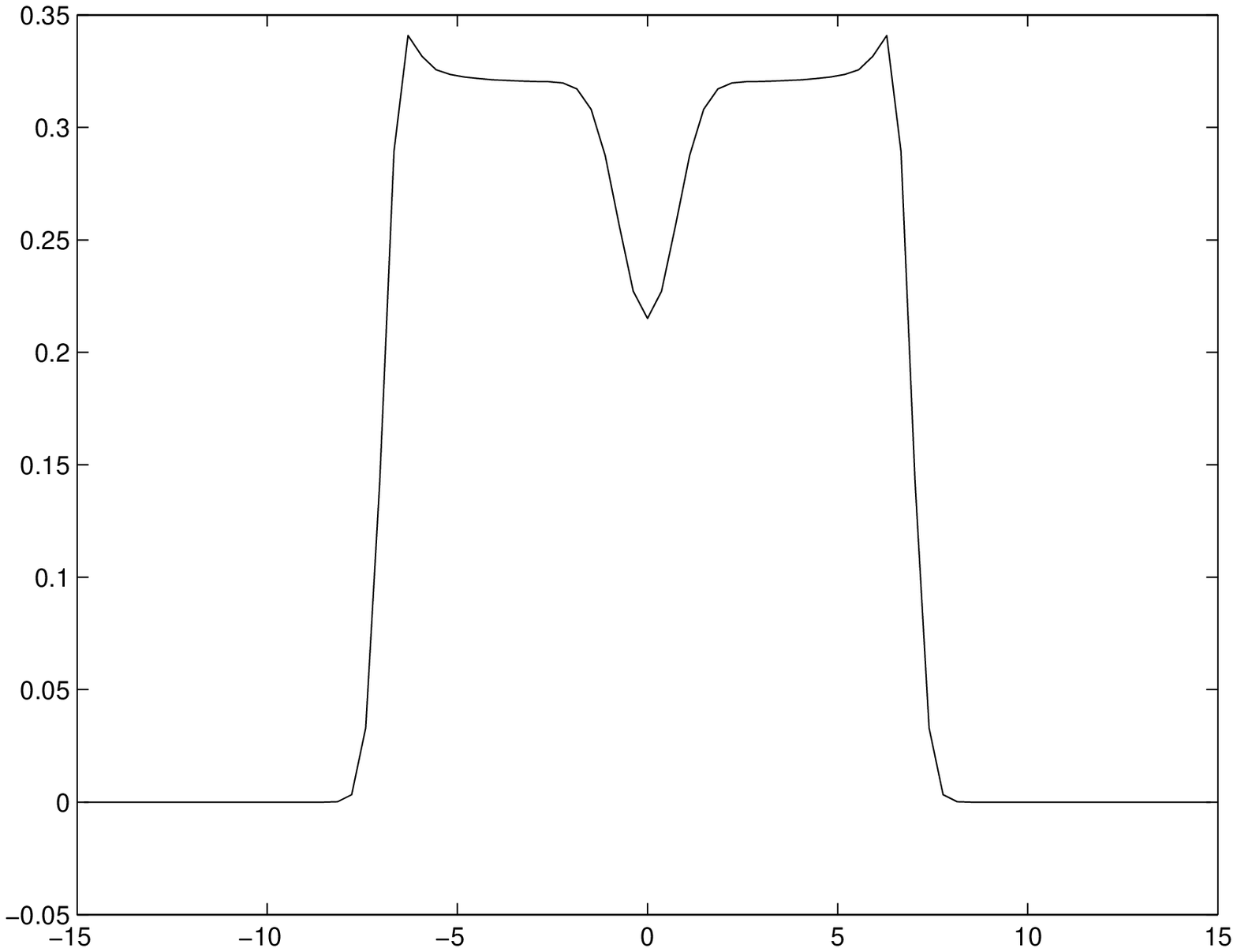}
}
\caption{
 \label{f:N50nu3qh3}
 The density distribution $\rho (\vec{x})$ and its profile given by (\ref{ourdef}) for (\ref{dropsolconqh}) for $N=50$, $\nu=q=1/3$.}.
\end{figure}

 The droplet solution (\ref{dropsolcon}) can readily be modified to describe a quasihole, \ie  a state where 
 the density close to the origin is depleted compared to the droplet state. Polychronakos found
 \be{dropsolconqh}
Z =\sqrt{q}|N\rangle\langle 0|+\sum_{n=1}^{N-1}\sqrt{n+q}|n-1\rangle\langle n|  \, ,
\ee   
where $0<q$ corresponding  to a shift in  the eigenvalues of the radius operator 
\be{X2}
R^2 = (X^1)^2+(X^2)^2=2\theta \sum_{n=0}^{N-1}(n+\frac{1}{2}+q)|n\rangle\langle n|  \, ,
\ee
with the amount $q$ relative to the original droplet. 
 By inserting (\ref{dropsolconqh}) into (\ref{ourdef}), we get the distribution $\rho(\vec{x})$ shown 
 in Fig.  \ref{f:N50nu3qh3}.

%%%%%%%%%%%%%%%%%%%%%%%%%%%%%%%%%%%%%%%%XXXXXXXXXXXXXXXXXXXXXXXXXXX

%%%%%%%%%%%%%%%%%%%%%%%%%%%%%%%%%%%%%XXXXXXXXXXXXXXXXXXXXXXXXXXXXX

\begin{figure}
\centerline{
\includegraphics[width=4cm]{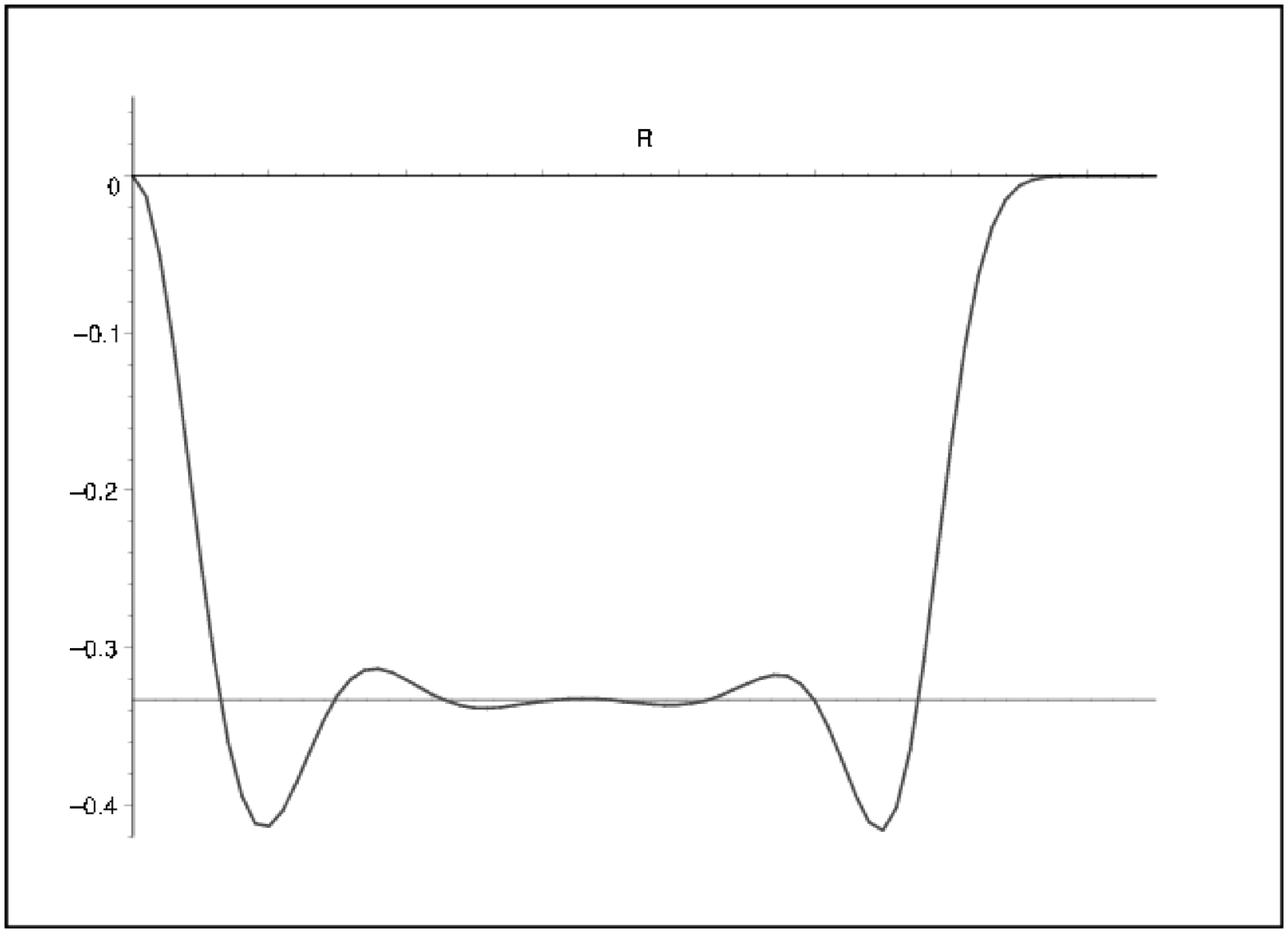} \hspace{2cm}
\includegraphics[width=4cm]{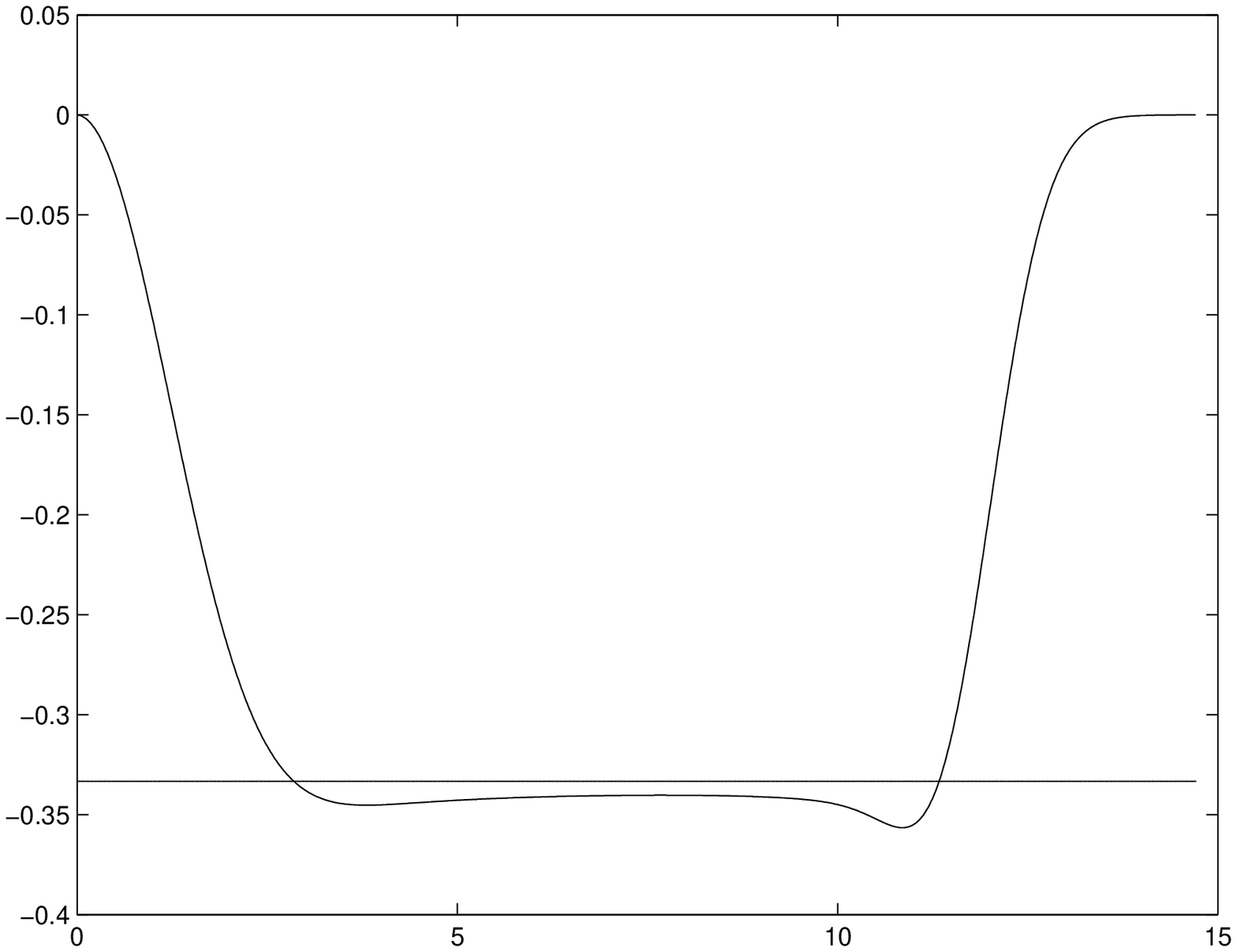}
}
\caption{
 \label{f:heidiqh}
 To the left, Monte Carlo calculation\cite{kjon} of the difference $Q_\mathrm{hole}(R)-Q_\mathrm{droplet}(R)$
 in the cumulative integrated charge between the Laughlin quasihole density distribution 
and the ground state density distribution  for $N=50$, $\nu=1/3$. The horizontal line corresponds to the charge difference $1/3$.
The radial distance is  in units of $\sqrt{2}\ell$ and the charge in units of $e$. To the right, the same distribution calculated from the matrix model. }
\end{figure}

Although there is a clear charge deficit at the origin, the matrix model does not reproduce the complete expulsion of the 
electrons characteristic of the Laughlin quasiholes. A more detailed comparison is made in Fig. \ref{f:heidiqh}, where we show the 
cumulative integrated charge $Q(R) = \pi \int_0^R dr^2\, \rho(r) $ for the Laughlin quasihole (left)\cite{kjon}  and the matrix model quasihole
 (\ref{dropsolconqh}) (right). We also calculated the 
 root mean square radius  for a quasihole of charge $\nu e$ in a state with filling fraction $\nu$ numerically
 with the results  $\sqrt{\av{r^2}}/\ell=2.6,\, 1.4,\, 1.1$  for $\nu=1,\frac{1}{3},\frac{1}{5} $ respectively. 
This can be  compared with the vortex solution of the mean field composite boson model (see \eg reference  \onlinecite{ezawa}), where the vortex has  $\av{r^2} = \nu \, 2.5 \ell^2$.

%%%%%%%%%%%%%%%%%%%%%%%%%%%%%%%%%%%%%%%%%%%%%%%%%%%%%%%%%%%%%%%%%%%%
 
%%%%%%%%%%%%%%%%%%%%%%%%%%%%%%%%%%%%%%%%%XXXXXXXXXXXXXXXXXXXXXXXXXX
 
\section{BPS solitons in the finite matrix model}

We already mentioned that the finite matrix model defined by \pref {mat} and \pref{fin} does 
not allow for quasielectron solutions. On the other hand, such solutions can be found if
the constraint is modified by hand. In the infinite matrix model we can  take
\be{modcon}
[ Z,Z^\dagger ] = 1 + q\ket 0 \bra 0\, ,
\ee
which describes a quasihole at the origin for $q>0$ and a quasielectron for $q<0$. To have dynamical 
quasielectrons a constraint of this type has to appear as one of the equations of motion. Such a construction, 
based on a noncommuative version of the Chern-Simons-Higgs model, was given by Bak \etal in reference 
\onlinecite{bak}. We first briefly review their work, and then show how to construct a 
corresponding finite matrix model. This will require both a modification of the action for the 
noncommutative scalar field, $\phi$, and  a coupling between $\phi$ and the boundary field $\Phi$.

In terms of the complex covariant position operators 
$Z=\frac{1}{\sqrt{2\theta}}(X^1+iX^2)$ and $Z^\dagger$, and the CS level number 
$ \kappa $, the noncommutative CS lagrangian \pref{mat} takes the form, 
\be{cs2}
L_\mathrm{CS} = 
\frac{i \kappa}{2} \tr \left(Z^\dagger \cD_0 Z - Z \cD_0 Z^\dagger\right) 
+\kappa \tr a_0,
\ee
to which Bak \etal  added a noncommutative $\phi^4$ lagrangian,  
\be{phif}
L_\mathrm{s} &=&  \tr \{ \phi^\dagger i \cD_0 \phi -
\frac{1}{2m} \cD_i\phi(\cD_i\phi)^\dagger+
\frac{\lambda}{2\theta} (\phi^\dagger\phi)^2 \}  \, .
\ee
Here $\phi$ is a matrix field in the fundamental representation,
\ie it transforms as $\phi \rightarrow U\phi $ under the gauge transformation $U$. 
The covariant derivatives are defined by  $D_\mu=\partial_\mu +i\hat{a}_\mu $ and act on $\phi $ as
\be{covder}
D_0 \phi &=& \partial_0 \phi +i \hat a_0 \phi  \nonumber \\
D_i\phi  &=& \frac i \theta \epsilon_{ij} [\hat{x}_j , \phi] + i\hat{a}_i \phi = 
\frac i \theta \epsilon_{ij}[X_j\phi - \phi \hat{x}_j]\, ,
\ee
where $\hat{x}_i$ are
noncommuting coordinates, $[\hat{x}_i,\hat{x}_j] = i\epsilon_{ij} \theta$. The corresponding derivatives are given by  $\partial_i= 
\frac i \theta \epsilon_{ij} [\hat{x}^j , \bullet]$, and  the matrices $X_i$, defining the actual state, are related
to the noncommutative gaugepotential via 
\be{xdev}
X_i = \hat{x}_i - \theta\epsilon_{ij}\hat{a}_j\, ,
\ee
 \ie  $\hat{a}_i$ parametrizes the deviation from the ground state solution 
$[X^1,X^2]=i\theta$.

Defining the current 
\be{topcur}
J_i &=& \frac{-i}{2m}
[(\cD_j\phi)\phi^\dagger -\phi(\cD_j\phi)^\dagger]   \, ,
\ee
the Hamiltonian can after some algebra be written as 
\be{bps}
H &=&  \tr \frac{1}{2m}(D_1\phi\pm iD_2\phi)(D_1\phi\pm iD_2\phi)^\dagger
 \pm \frac{\epsilon_{ij}}{2}\tr D_iJ_j \pm 
\frac{1}{2m\theta} \tr \left([Z,Z^\dagger]- 1 \mp  
\lambda m\phi\phi^\dagger \right) \phi \phi^\dagger. 
\ee
Assuming the solution to be regular enough for the covariant derivative of the current to integrate to zero, and 
taking  $\lambda m \kappa = \pm 1 $ so that  the last term will vanish because of  the constraint
\be{conwphi}
[Z,Z^\dagger]= 1-\frac{1}{\kappa} \phi\phi^\dagger ,   
\ee
the Hamiltonian reduces to the first term which is quadratic and equals zero for $(D_1\phi \pm iD_2\phi)=0$.
Thus this choice of parameters corresponds to the theory being of the Bogomol'nyi-Prasad-Sommerfield (BPS) form \cite{jack,bak}.
The complete set of BPS equations is given by, 
\be{bpseq}
\hat a_0     &=&\frac{1}{2\kappa\theta}\phi\phi^\dagger  \\
D_1\phi \pm iD_2\phi   & =& 0 \label{bps1} \nonumber  \\
{[}Z,Z^\dagger {]}   &=& 1-\frac{1}{\kappa} \phi\phi^\dagger \nonumber,   
\ee
and it is easy to check that any solution of these is also 
 a solution to the full time-independent equations of motion corresponding to $L_{CS} + L_s$.

We now turn to finite matrices. 
Because of the coupling $\sim \tr (\hat a_0\phi\phi^\dagger)$ the Gauss law constraint became (\ref{conwphi}), which can  be satisfied 
 by finite matrices. Of course,  
it is then no longer possible to  have 
$[\hat{x}^1,\hat{x}^2] = i\theta [z,z^\dagger]=i\theta $,
where $z=\frac{1}{\sqrt{2\theta}}(\hat{x}^1+i\hat{x}^2)$.
 Instead  we let
\be{defM}
[\hat{z},\hat{z}^\dagger]=M= 1 - N\ket {N-1} \bra {N-1} \, ,
\ee
with 
\be{defz}
\hat{z}=
\sum_{n=1}^{N-1}\sqrt{n}\ket{n-1}\bra{n}\, .
\ee
Now the Hamiltonian can no longer be written on  BPS form (\ref{bps}), 
but  by adding the term,
\be{newterm}
L_M= \mp \frac{1}{2m\theta} \tr (+\phi M\phi^\dagger- \phi\phi^\dagger) \, ,
\ee
it is a matter of algebraic manipulations to 
show that the Hamiltonian corresponding to $L = L_{CS} + L_s +L_M$ is again of the BPS type.

It is not hard to verify that the model we just defined has droplet
solutions, and topological solitons of the type found by Bak \etal\cite{bak}. 
There are however no quasihole solutions. This can be remedied by 
also adding a Polychronakos type boundary field, which has the additional advantage
that the sector where the scalar field is not excited becomes 
 identical to the original finite QH matrix model. Our final lagrangian now  reads
\be{totmod}
L = L_{CS} + L_s + L_M +  L_\Phi
\ee
where the boundary lagrangian is given by,
\be{bflag}
L_\Phi =   
\Phi^\dagger i\cD_0 \Phi - 
\frac{\lambda }{2\theta}\Phi^\dagger \phi \phi^\dagger \Phi \,       
\ee
yielding the Gauss law constraint
\be{fincon}
 {[}Z,Z^\dagger{]} &=& 1-\frac{1}{\kappa}\phi\phi^\dagger
-\frac{1}{\kappa}\Phi\Phi^\dagger   \, .
\ee
The last  term in (\ref{bflag}) was added to allow the Hamiltoninan to have a  BPS form almost identical  to \pref{bps}, but with the BPS equation \pref{conwphi} replaced by \pref{fincon}.
The remaining BPS equations are unchanged.
This completes the derivation of the extended finite QH matrix model,
which is  a finite matrix version of the conformal Chern-Simons-Higgs model
introduced by Jackiw and Pi\cite{jack}.

We now turn to a discussion of the solutions of this model. First note that 
 all solutions discussed in section II, \ie the isolated particles \pref{symsolcon}, 
the droplet \pref{dropsolcon}, and the quasihole \pref{dropsolconqh}, can all 
be taken over unchanged if we set $\phi = 0$. For non-zero $\phi$ we will have two new
types of solutions corresponding to waves and solitons.  The latter, which 
will provide the basic building block for the quasielectrons, are the most interesting,
but we first briefly discuss the former.

\subsection{Collective modes} 
For our model to give a 
realistic description of the QH system it is important that the collective 
wave-like solutions in the bulk are gapped. This is certainly expected from the analogy with the continuum model, but should nevertheless be established 
in the matrix model context. Let us first consider the case of a 
constant density of $\phi$ particles, $\rhot$, (not to be confused with the 
constant density of electrons $\bar\rho=\frac{1}{2\pi\theta}$ represented 
by the solution $\hat{Z}=\hat{z}, \phi = 0$) in the infinite matrix model 
of Bak \etal \cite{bak}. 
The mean field solution that we want to expand about is given as an 
expansion in the density $\rhot$
\be{meanfield}
Z &=& \left(1-\frac{\rhot}{2\kappa}\right)z+{\cal O}(\rhot^2)
\nonumber\\
\phi &=& \sqrt{\rhot} \left(\hat{1}+{\cal O}(\rhot)\right)
\\
a_0 &=& -\frac{\lambda}{2\theta}\rhot +{\cal O}(\rhot^2)
\nonumber
\ee
which solves the full equations of motion to first order in $\rhot$.

We then expand  \pref{cs2} and \pref{phif} to 
quadratic order around the mean field solution, and then use the polar 
decomposition $\phi = UP$ which is valid for an arbitrary square matrix. 
Here $U$ is a unitary, and $P$ is a positive semi-definite hermitian matrix 
\cite{metha}. By a $U(N)$ gauge transformation, we can now remove 
the $U(N)$ phase $U$ from the field $\phi$. As a result we find 
that the kinetic term of what remains of $\phi$ becomes a total 
derivative and this field thus becomes a Lagrange multiplier enforcing a 
constraint relating the fluctuations of the gauge field and $\phi$. The 
result is that we have moved 
the entire dynamics from the scalar field to the gauge field $Z$. This is 
the noncommutative version of going to unitary gauge in Ginzburg-Landau Chern-Simons (GLCS) theory. The 
resulting lagrangian reads,  
\be{exp}
L_{CS} +  L_{\phi} = 
i\kappa \tr a^\dagger \dot a -\frac {\tilde\rho} {2m}  \tr  (a a^\dagger + a^\dagger a) + 
\dots \, ,
\ee
in terms of $\hat{a}\equiv i(\hat{z}-Z)=\sqrt{\theta/2}(\hat{a}_1+i\hat{a}_2)$, with $\hat{a}_i$ given by \pref{xdev}.
The dots indicate 
commutator terms corresponding to spatial derivatives, as well as 
potential terms and terms of higher order in $\rhot$. There is also a 
constraint equation that relate density fluctuations to the noncommutative 
gauge field. Just as in the commutative case, the lagrangian \pref{exp} has the form of a 
harmonic oscillator, and consequently exhibits a gap at $\omega_c = 
\rhot/\kappa m $. This has the natural interpretation as the Kohn mode 
at the cyclotron frequency of the $\phi$ particles.

For solutions with vanishing background density $\rhot$ - the simplest 
case being that of a soliton considered below -  there will be zero modes 
corresponding to translations. In the full model \pref{totmod} there will 
also be gapless edge modes. We have not analyzed these more complicated 
cases, but we think that the above demonstration of the similarity 
between the commutative and noncommutative models strongly suggests that 
the latter will not develop any gapless modes not found in the 
former.

\subsection{Solitons and quasielectrons}

In reference \onlinecite{bak}, Bak \etal found noncommutative counterparts of the self dual vortex solutions
due to Jackiw and Pi. These correspond to a quantized flux, and, as will be clear from the 
explicit expressions given below, they carry unit electric charge. Since flux is quantized, one cannot 
have fractionally charged quasielectrons in the model by Bak {\em et.al.}, but in our finite matrix model
there is a natural construction in terms of a soliton combined with a quasihole. 

\subsubsection {The charge $-1$ soliton}

\begin{figure}
 \begin{center}
     \includegraphics[scale=0.666]{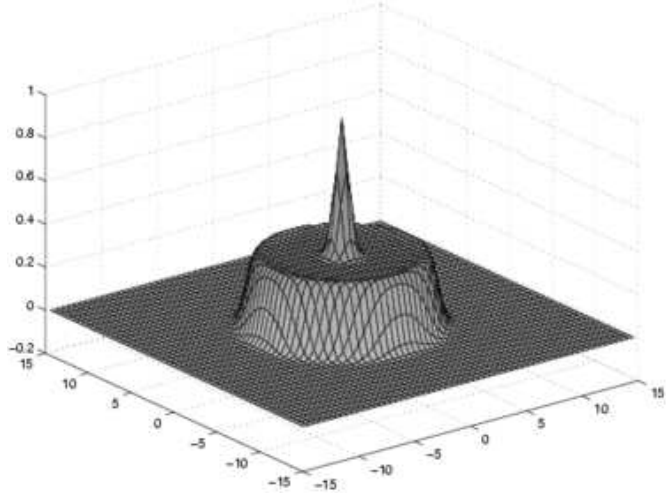}
 \end{center}
 \caption{The density distribution of a soliton of charge $-1$ for $N=50$, $\nu=1$. }
 \label{f:sem50rhoxys}
 \end{figure}

Using \pref{covder} one can derive
\be{bpsZz}
D_1\phi  - iD_2\phi = -  \sqrt{\frac{2}{\theta}}(Z^\dagger\phi -\phi \hat{z}^\dagger )\, , 
\ee
and we will take the sign in (\ref{bps})  such that  this is one of the BPS equations \pref{bpseq}. 
A  soliton of charge $-1$ centered at the origin is now given by the following expression, 
\be{soliton}
Z &=& \sum_{n=2}^{N-1}\sqrt{n-1}\ket{n-1}\bra{n} \nonumber  \\ 
\Phi&=&  \sqrt{(N-1)\kappa} \ket{N-1}   \\ 
\phi&=& \sqrt{\kappa} \ket{0}\bra{0} \, . \nonumber
\ee
It is easy to verify by direct substitution that this indeed is a solution to the BPS equations corresponding
to the full QHMM \pref{totmod}.
In Fig.~\ref{f:sem50rhoxys} we show a density plot of this solution for $N=50$.

%%%%%%%%%%%%%%%%%%%%%%%%%%%%%%%%%%%%%%%XXXX

%%%%%%%%%%%%%%%%%%%%%%%%%%%%%%%%%%%%%%%%XXXXXX

\subsubsection{The quasielectron}

In analogy with the soliton solution \pref{soliton}, we can now try to construct a fractional quasi-electron  based
on the Ansatz,
$
Z =\sum_{n=1}^{N-1}\sqrt{n+q}\ket{n-1}\bra{n} 
$
with $-1<q<0$. Note, however, that  this implies $Z \ket{0} \neq 0$ 
so the only option seems to be 
$\phi=\sqrt{(N+q)\kappa} \ket{N-1}\bra{0}$
and 
$\Phi=\sqrt{-q\kappa} \ket{0}$.  Such a solution would however not reduce to the 
ground-state (\ref{dropsolcon}) as $q\rightarrow 0$. 

\begin{figure}
\centerline{
     \includegraphics[scale=0.666]{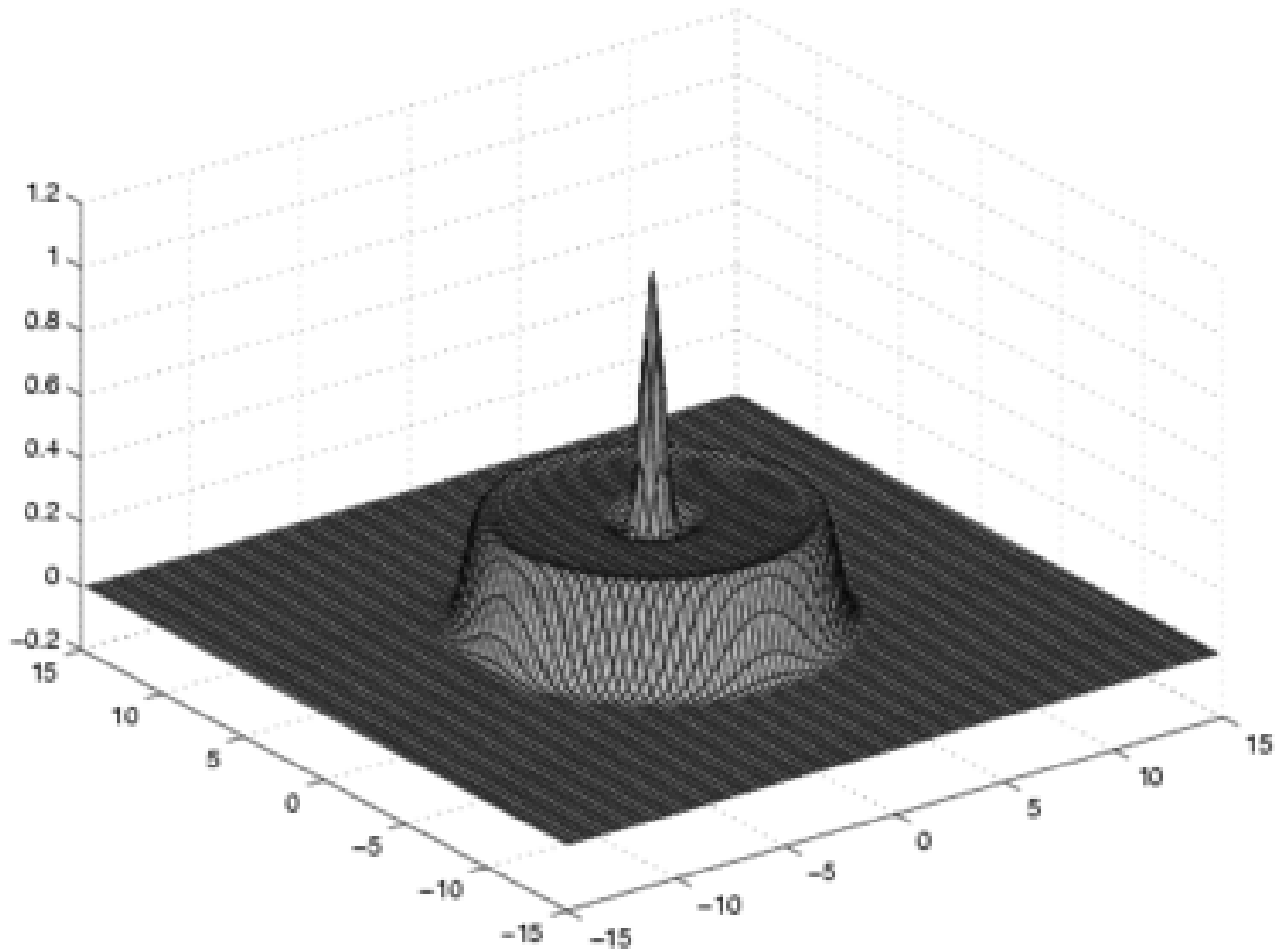}
 \includegraphics[scale=0.2]{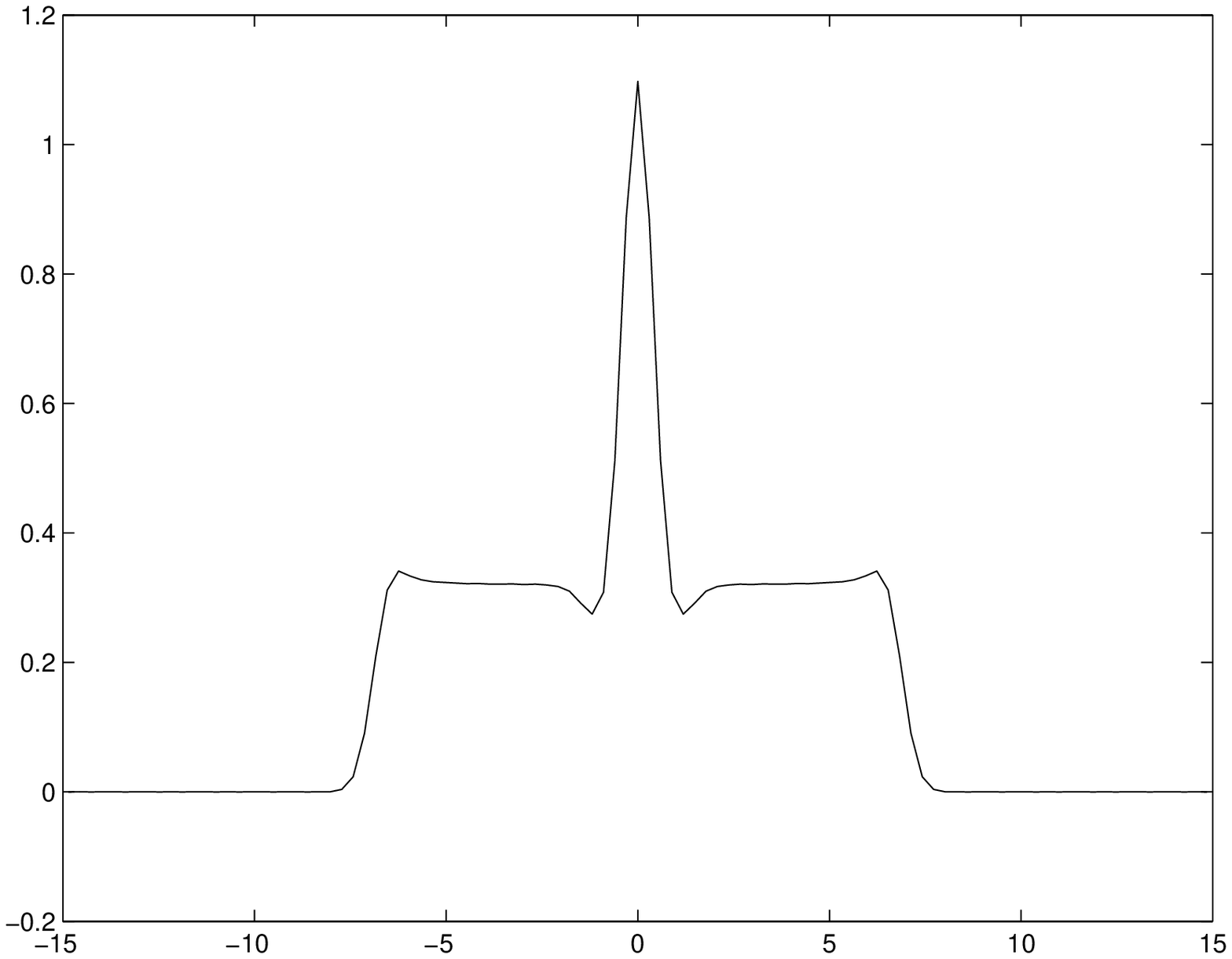}
}
 \caption{The density distribution of 
 a  charge $-\frac{1}{3} $ quasielectron 
composed of  a charge $+\frac{2}{3} $ quasihole on top of  a charge -1 soliton 
for $N=50$, $\nu=1/3$. }
 \label{f:N50nu3qe3}
 \end{figure}

\begin{figure}
\centerline{    
 \includegraphics[width=4cm]{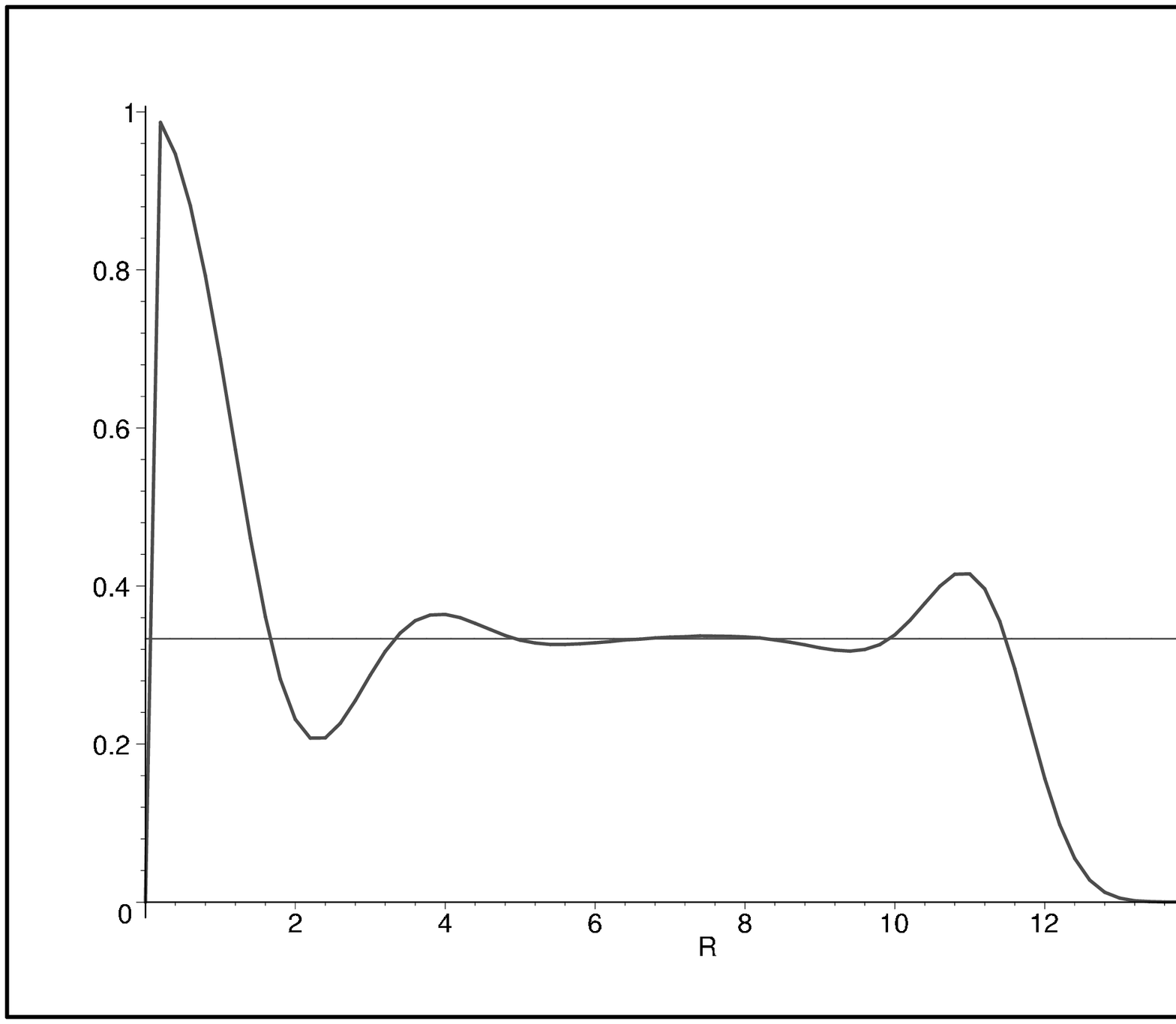}
\hspace{2cm}
\includegraphics[width=4cm]{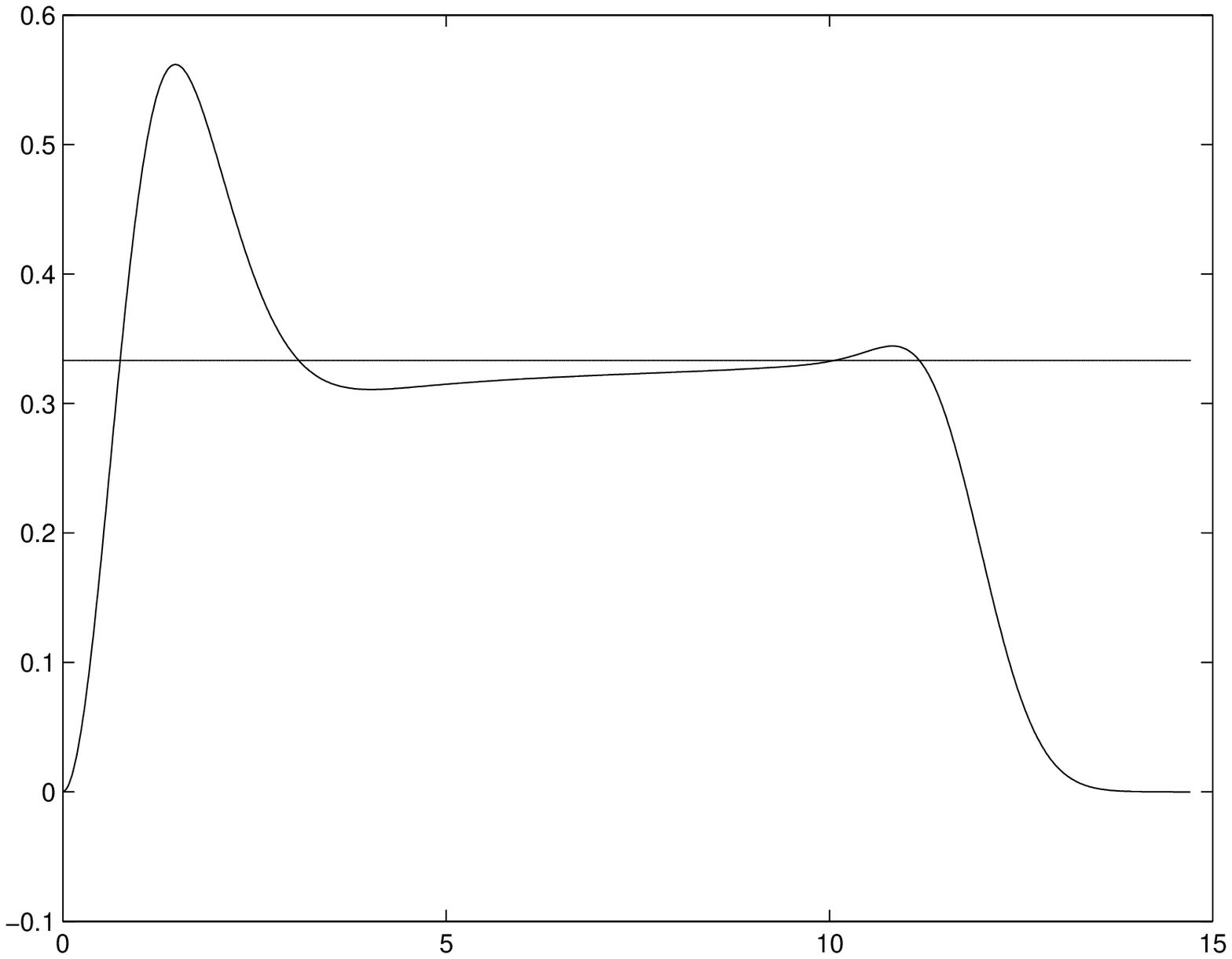}
}
 \caption{The same  plots as in Fig. \ref{f:heidiqh}, but  for a Jain quasielectron.\cite{kjon} Note the characteristic dip at the foot to the right of the quasielectron.\label{f:heidijqe}}
\end{figure}

An obvious alternative construction, alluded to above, 
is to add a quasihole to the soliton, \ie  to combine the solutions
\pref{soliton} and \pref{dropsolconqh}. This amounts to finding  a $q$-dependent 
modification of $Z$ for which the constraint is still $q$-independent. The solution is
\be{qp}
Z &=&  \sqrt{q}\ket{N-1}\bra{1}+ \sum_{n=1}^{N-1}\sqrt{n-1}\ket{n-1}\bra{n}   \nonumber  \\ 
\Phi&=&  \sqrt{N\kappa}\ket{N-1}  \\
\phi&=& \sqrt{\kappa} \ket{0}\bra{0} \nonumber \, , 
\ee
for which we still have $Z \ket{0}=0$ and hence $\bar{\cD} \phi=0$. 
Again the BPS equations can be verified by direct substitution. 
In Fig.\ref{f:N50nu3qe3} we show  for $N=50$, $\nu=1/3$ a $q=2/3$ hole sitting 
on the top of a  soliton of charge $-1$ to produce  a charge $-1/3$ quasielectron. The small dip at the foot of the peak has also been seen in numerical 
studies of QH wave functions\cite{kjon}, see Fig. \ref{f:heidijqe}. 

%%%%%%%%%%%%%%%%%%%%%%%%%%%%%%%%%%XXXXXXXXXXXXXX

%%%%%%%%%%%%%%%%%%%%%%%%%%%%XXXXXXXXXXXXXXXXXXXXXXXXXXXXXX

%%%%%%%%%%%%%%%%%%%%%%%%%%%%%%XXXXXXXXXXXXXXXXXXXXXXXXXXXX

%%%%%%%%%%%%%%%%%%%%%%%%%%%%XXXXXXXXXXXXXXXXXXXXXXXXXXXX

Note that this solution does not reduce to the ground state as $q\rightarrow 1$. This is not necessarily a drawback, since it might be interpreted as the limiting case of 
a small exciton, \ie an overlapping state of an electron and a hole. 
Finally we should mention that we have not investigated the stability of our quasielectron solution, so we cannot be
sure that it will not decay into a soliton and a quasihole. Although such a calculation amounts to a straightforward 
small oscillation analysis, it is algebraically complicated, and also of limited interest since we would in any case have  
a stabilizing Coulomb interaction in a more detailed model.

\section{Summary and discussion}
 To summarize, we have argued for a particular expression for the charge density in the {\em classical} QH matrix models, and shown that with this definition, the various solutions  corresponding to separated particles, droplets and quasiholes are 
 reproduced in reasonable agreement with standard treatments based on wave functions and GLCS mean field solutions. We furthermore extended the model to incorporate densities higher than that of the groundstate, and found quasielectron solutions. Again the profiles were in good agreement with those found from explicit wave functions. 
 
In this connection it is fair to ask what has been gained by the classical QH matrix model as compared
to the usual mean field CS theories, so we shall now briefly contrast these approaches. 
That the two theories are closely connected is clear from Susskind's original formulation of the QH matrix
model as a noncommutative CS theory described by,
 \be{nceff}
 {\cal L} = \frac 1 {4\pi\nu} \epsilon^{\mu\nu\lambda}\left(
 a_{\mu}\star\partial_{\nu}a_{\lambda}  + 
 \frac {2i} 3 a_\mu\star a_\nu \star a_\lambda  \right)  \, ,
 \ee
 where the Moyal star product $\star$ is defined with a noncommutative parameter $\theta = (2\pi\bar\rho)^{-1}$. 
This is a purely topological theory, consistent with the infinite matrix model having a unique state
with a constant density $\bar\rho$. The corresponding commutative CS theory is also topological and 
has a unique state on an infinite plane.  Here we should note that this commutative effective CS theory can be derived from the GLCS theory by expanding around a mean field.\footnote{ 
%%%%%%%%%%%%%%%%%%%%%%%%%%FFFFFFFFFFFFFFFFFFFFFFFFFFFFFFFFFFFFFFF
Since the 
quantum GLCS theory is a direct rewriting of the microscopic theory (using a singular gauge transformation), it
is an interesting open question whether the noncommutative action\cite{zhang92} could be obtained by a more sophisticated mean field approach. }
%%%%%%%%%%%%%%%%%%%%%%%%%%%FFFFFFFFFFFFFFFFFFFFFFFFFFFFFFFFFFFFFFF

Quasielectrons and quasiholes can be introduced by hand in the infinite matrix model by changing the 
constraint. In the CS theory this correspoinds to adding delta function sources. Here we see the first 
advantage of the matrix model in that it gives a size $\sim \theta$  to the quasi particles.\footnote{
%%%%%%%%%%%%%%%%%%FFFFFFFFFFFFFFFFFFFFFFFFFFFFFFFFFF
By introducing further terms in the expansion around the mean field in the commutative CS theory, one
generates terms $\sim b^2$, where $b$ is the CS magnetic field. Such a term will give a size to the
quasiparticle, as discussed in \eg reference \onlinecite{artz}. 
 }
%%%%%%%%%%%%%%%%%%%%FFFFFFFFFFFFFFFFFFFFFFFFFFFFFFF

Adding the boundary field $\Phi$ to the matrix model allows for a plethora of states not described by the
usual CS approach. Defining the latter on a manifold with a boundary gives edge degrees of freedom corresponding to
chiral Luttinger liquids, but there are no excitations inside the bulk nor outside the droplet. The basic reason is
that the edge excitations in the CS theory can be understood as hydrodynamic modes of an incompressible 
liquid, while the matrix model allows for density fluctuations in the fluid itself. From this it is also clear that no
questions regarding density profiles or effective sizes of quasiparticles can be addressed in the framework of pure
CS theory. 

There is an asymmetry between quasielectrons and quasiholes in the matrix model, since there is a maximal density  given by the noncommutative parameter $\theta$.  This was the basic reason that forced us to introduce a 
 new field, $\phi$, to describe quasielectrons, while quasiholes were present already in the model based on
 only a CS field $X^i$ and the boundary field $\Phi$ needed to ''absorb'' the anomaly. 
 Such an asymmetry is present also in other descriptions of the QH effect. For instance, in 
  the wave function approach Laughlin's quasihole wave function is essentially unique, while there are
  several quite different approaches to the quasielectron state\cite{kjon}.  The introduction of a new field 
  raises questions about the correct counting of degrees of freedom. The finite matrix model without
  any extra field describes $N$ particles, but with a phase space repulsion giving a maximum density $\sim 1/\theta$. 
 As we have shown, the extra field relaxes the maximum density constraint in a way consistent with QH phenomenology, 
 but one might worry that we have at the same time introduced additional unphysical (gapped) excitations in 
 the high energy part of the spectrum. We have not investigated this problem any further.

To summarize, there are some aspects of QH physics that is more easily described in conventional CS framework, notably the classification of abelian QH liquids developed by Wen\cite{zee}. The QH matrix model, on the 
other hand, allows for a more detailed analysis of density profiles and a dynamical description of quasielectrons 
and quasiholes. 

There are several detailed questions left open concerning the details of the classical QHMM, and the quantum 
theory is to a great extent unexplored territory. If we might venture a guess, we would however say that if the 
noncommutative approach to QH physics is to provide any essential new physical insights one has either to 
find ways to generalize the quantum models - with the aim of  understanding the hierarchy and/or the Jain states - or to find some quantitative use for the classical description. The mere fact that a classical model can do so well in describing 
a strongly interacting system in the extreme quantum regime is in itself intriguing, and it might  be quite 
interesting to extend the model to include disorder and study possible phase transitions. 
 
 \vskip 3mm \noi {\bf Acknowledgment:} We thank Alexis Polychronakos for interesting discussions and helpful comments
 on the manuscript. THH and AK were supported by the Swedish Research Council and the work of RvU was
supported by the Czech ministry of education under contract no. 143100006.

 %%%%%%%%%%%%%%%%%%%%%%%%%%%%%%%%%

\appendix
\section{More on  density operators}
In this appendix we first demonstrate that the definition of $\rho(\vec x)$ implied by \pref{ourdef}  gives negative 
values for certain configurations satisfying the constraint \pref{ptvang}. We then give two alternative definitions of 
 $\rho(\vec x)$ which are positive, but have other difficulties. 

\subsection{The Weyl-ordered density is not positive}

\begin{figure}
\begin{center}
\includegraphics[width=5cm]{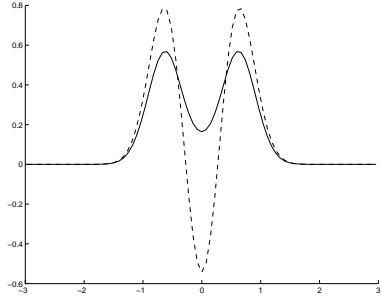}
\end{center}
\caption{
 \label{f:psdfrhfcomps1}
 Density profiles of the droplet solution \pref{dropsolcon} for $N=2$, $\nu=1/5$.  The definition (\ref{ourdef})
 is given by the broken line and the positive definite definition 
(\ref{newdef}) by the full line. Both definitions  give a 
circularly symmetric distribution.
 The density \pref{newdef} integrates to a total particle number of about 1.66.}
\end{figure}

For small electron numbers and small filling fractions one can find many solutions for where the density becomes negative. In some cases it just about becomes negative, but for other solutions the violation of positivity is big, as can be seen in Fig. \ref{f:psdfrhfcomps1}. 
%%
%%%%%%%%%%%%%%%%%%%%%%%%%%%%%%%%XXXXXXXXXXXXXXXXXXXXXXXXXXXXXXXXX

%%%%%%%%%%%%%%%%%%%%%%%%%%%XXXXXXXXXXXXXXXXXXXXXXXXXXXXXXXXX

That the density is sometimes negative can, for special cases, also be established analytically to lowest order in $\theta$.

\subsection{Alternative density operators}
We now give two alternative definitions of the density operator that are both non-negative. 
The starting point is the  non-relativistic density operator for $N$ point particles,
\be{denx}
\rho(\vx) = \sum_{n=1}^N \delta (\vx - \vxi n) \, .
\ee
If $\vx_n$ are taken as quantum operators, this is also the first quantized density operator
in the $\vx$ representation. The operator \pref{denx} is by construction non-negative, since it 
is a sum of positive operators. In momentum space, the operator \pref{denx} takes the form
\be{denp}
\rho (\vk)  = \sum_{n=1}^N e^{-i \vk\cdot\vxi n } \, .
\ee

It is well known that quantum mechanical particles in the lowest Landau level are described by the following
density operator, 
\be{ldenp}
\rho_L (\vk)  = \sum_{n=1}^N e^{-\frac i 2 \bar k z_n }e^{-\frac i 2  k \bar z _n}
\ee
with $z_n=x_n^1+i x_n^2$ and $[z_m,\bar{z}_n] = 2\ell\delta_{mn}$. In $\vx$ space this becomes,
\be{ldenx}
\rho_L (\vx)  = \frac 1 {2\pi\ell^2} \sum_{n=1}^N e^{-\frac { (\vx - \vxi n)^2 } {2\ell^2} }  
= \sum_{n=1}^N \delta_\ell(\vx - \vxi n) \, , 
\ee
where $\delta_\ell$ can be thought of as a regularized delta function. Again the operator 
\pref{ldenx} is positive by construction.  

With these preliminaries, we now present two possible definitions of $\rho(\vx)$ in the finite matrix model that 
are  manifestly  positive. The most obvious idea is to try to extract a set of $N$ particle positions, $\vxi n$ from the 
matrices $X^i$ and simply plug these into a formula of the type \pref{ldenx}. In this case we are of course free to 
use any positive definite profile function for the particles, but by choosing exactly \pref{ldenx} we ensure that
the profile of a single particle in the matrix model is identical to that of an electron in the lowest Landau level. 
The problem of defining coordinates in the QH matrix model was discussed in a paper by Karabali and Sakita\cite{kara01}.
They showed that taking the eigenvalues of the complex matrix $Z$ as particle positions,
\footnote{ 
%%%%%%%%%%%%%%%%%%%%%%%%%%%%%%%%%%%%
All but a set of measure zero of the complex  matrices $Z$ can  be diagonalized as $Z = XEX^{-1}$. }
%%%%%%%%%%%%%%%%%%%%%%%%%%%%%%%%%%%%%
correctly reproduced the low momentum part of the Laughlin wave function, while the short distance part was 
distorted - the characteristic $|z_i - z_j|^{2\kappa}$ behaviour of the two particel correlation was softened to a lower power.  We would thus expect that a density operator defined by \pref{ldenx} and the coordinates proposed in 
reference \onlinecite{kara01} in spite of being positive, would have difficulties in describing the profiles studied in this paper, which vary rapidly on the order of a magnetic length. Since the construction is very indirect, we also 
do not have any closed expression for the density and current similar to \pref{charge}.

Another possibility is based on expressing the operator \pref{ldenx} as a square of an operator, thus
making the positivity manifest:
\be{sqxdel}
\rho_L (\vx)  = (\rho_L (\vx)^\half)^2\, ,
\ee
where
\be{approx}
\rho_L (\vx)^\half 
 \approx \frac 1 {\sqrt{2\pi\ell^2}}
\sum_{n=1}^N e^{-\frac { (\vx - \vxi n)^2 } {4\ell^2} }  \approx   \sum_{n=1}^N \delta_\ell^\half (\vx - \vxi n) \, , 
\ee
which is a good approximation when the particles are far apart. 
Going  to Fourier space, where the square of the distribution become a convolution integral, we 
are led to the following proposal for the density operator,
\be{newdef}
\rho_{pos} (\vp) = \frac{2 \ell^2}{ \pi} \int \mathrm{d}^2k\, e^{-\ell^2[(\vp -\vk)^2 + k^2] }
\tr [e^{-i(\vp -\vk)\cdot \vec X ]} \tr [e^{-i \vk \cdot \vec X } ] \, . 
\ee
The corresponding $\rho_{pos}(\vx)$ is positive by construction, and it is easy to show that for 
widely separated particles, where the matrices become almost diagonal, the profile reproduces 
the one given by \pref{sqxdel} and \pref{approx}. When the particles come closer this is no longer
true. Figure \ref{f:psdfrhfcomps1} shows the droplet solution with 
the old definition \pref{ourdef} shown by a broken line, and the definition  (\ref{newdef}) by a solid  line.  
Other examples, like for particles further apart, show again that if the density becomes  negative, (\ref{newdef}) repairs that. The basic problem of that definition is that it is not normalized, \ie $\int d^2x\, \rho_{pos}(\vx)\neq N $.This can of course be remedied by a renormalization, but difficulties remain.

 Note that the definition \pref{newdef}
is not in the general class \pref{charge} since it involves the product of two traces rather than a single trace over
 a matrix kernel. This in particular means that our construction of a conserved current is no longer valid, but more importantly, that  Pandora's box is opened - why should we restrict ourselves to the product of two 
 traces? Why not several, or perhaps even an infinite series? 

 In summary, we have given alternative constructions of the density operator which are manifestly non-negative. 
There are however other difficulties related to these proposals, and we have no reason to believe that they 
would provide a better description than \pref{ourdef} that we used in the main text of the paper.

 \end{document}